\PassOptionsToPackage{table}{xcolor}
\documentclass[AMA,STIX2COL,Linenumbersoff]{MRM}
\articletype{Research Article}%
\usepackage{amsfonts,ulem,lscape}
\usepackage{color, rotating}
\definecolor{Gallery}{rgb}{0.937,0.937,0.937}
\DeclareMathOperator*{\argmin}{argmin}
\received{XX}
\revised{XX}
\accepted{XX}
\doi{, Word Count: $\sim$5000}
\definecolor{Green}{RGB}{34,139,34}
\begin{document}

\title{Coil Sketching for computationally-efficient MR iterative reconstruction}

\author[1,2]{Julio A. Oscanoa}{\orcid{0000-0002-2847-8896}}

\author[3]{Frank Ong}{\orcid{0000-0002-9789-8683}}

\author[4]{Siddharth S. Iyer}{\orcid{0000-0002-6432-0850}}

\author[2]{Zhitao Li}{\orcid{0000-0003-3462-6549}}

\author[3]{Christopher M. Sandino}{\orcid{0000-0002-8360-0153}}

\author[3]{Batu Ozturkler}{}

\author[2]{Daniel B. Ennis}{\orcid{0000-0001-7435-1311}}

\author[3]{Mert Pilanci}{}

\author[2]{Shreyas S. Vasanawala}{\orcid{0000-0002-1999-6595}}

\authormark{OSCANOA \textsc{et al}}

\address[1]{\orgdiv{Department of Bioengineering}, \orgname{Stanford University}, \orgaddress{\state{California}, \country{USA}}}

\address[2]{\orgdiv{Department of Radiology}, \orgname{Stanford University}, \orgaddress{\state{California}, \country{USA}}}

\address[3]{\orgdiv{Department of Electrical Engineering}, \orgname{Stanford University}, \orgaddress{\state{California}, \country{USA}}}

\address[4]{\orgdiv{Department of Electrical Engineering and Computer Science}, \orgname{Massachussets Institute of Technology}, \orgaddress{\state{Massachussets}, \country{USA}}}

\corres{Julio A. Oscanoa. Department of Radiology, Stanford University, 1201 Welch Road, Stanford, CA 94305, USA. \email{joscanoa@stanford.edu}}

% \presentaddress{This is sample for present address text this is sample for present address text}

\finfo{National Institutes of Health : R01-EB026136 : R01-EB009690 : R01-EB019241.\\ National Science Foundation : ECCS-2037304 : DMS-2134248 : CCF-2236829. \\U.S. Army Research Office Early Career Award : W911NF-21-1-0242. \\Stanford Precourt Institute. \\ ACCESS—AI Chip Center for Emerging Smart Systems through InnoHK, Hong Kong, SAR.
 }

\abstract{
\vspace{-0.2cm}
\normalsize
\section{Purpose} Parallel imaging and compressed sensing reconstructions of large MRI datasets often have a prohibitive computational cost that bottlenecks clinical deployment, especially for 3D non-Cartesian acquisitions. One common approach is to reduce the number of coil channels actively used during reconstruction as in coil compression. While effective for Cartesian imaging, coil compression inherently loses signal energy, producing shading artifacts that compromise image quality for 3D non-Cartesian imaging. We propose coil sketching, a general and versatile method for computationally-efficient iterative MR image reconstruction.
\section{Theory and Methods} We based our method on randomized sketching algorithms, a type of large-scale optimization algorithms well established in the fields of machine learning and big data analysis. We adapt the sketching theory to the MRI reconstruction problem via a structured sketching matrix that, similar to coil compression, considers high-energy virtual coils obtained from principal component analysis. But, unlike coil compression, it also considers random linear combinations of the remaining low-energy coils, effectively leveraging information from all coils.
\section{Results}  First, we performed ablation experiments to validate the sketching matrix design on both Cartesian and non-Cartesian datasets. The resulting design yielded both improved computational efficiency and preserved signal-to-noise ratio (SNR) as measured by the inverse g-factor. Then, we verified the efficacy of our approach on high-dimensional non-Cartesian 3D cones datasets, where coil sketching yielded up to three-fold faster reconstructions with equivalent image quality.
\section{Conclusion} Coil sketching is a general and versatile reconstruction framework for computationally fast and memory-efficient reconstruction. 
}

\jnlcitation{\cname{%
\author{Oscanoa JA}, 
\author{Ong F}, 
\author{Iyer SS}, 
\author{Li Z}, 
\author{Sandino C},
\author{Ozturkler B},
\author{Ennis DB},
\author{Pilanci M}, and
\author{Vasanawala SS}
} (\cyear{2023}), 
\ctitle{Coil Sketching for computationally-efficient MR iterative reconstruction}, \cjournal{Magn. Reson. Med.}, \cvol{XXXX;XX:X--X}.}

\maketitle
% \footnotetext{\textbf{Abbreviations:}~\hbox{ANA,~anti-nuclear~antibodies;~APC,~antigen-}{\hfill\break}presenting~cells; IRF, interferon regulatory factor}
\section{Introduction}\label{sec1}
The combination of parallel imaging \cite{pruessmann1999sense, griswold2002generalized, sodickson1997simultaneous} and compressed sensing \cite{lustig2007sparse} (PICS) is one of the most successful techniques to shorten scanning times in Magnetic Resonance Imaging (MRI) \cite{king2008combining,tariq2013venous,zhang2014clinical,murphy2012fast}. However, removal of undersampling artifacts in high-dimensional MRI datasets requires image reconstruction algorithms which have long computation times and considerable memory usage. These drawbacks have hindered clinical deployment of high-dimensional applications such as 3D non-Cartesian acquisitions \cite{gurney2006design}, 4D-Flow \cite{sandino2017accelerated}, Dynamic Contrast-Enhanced MRI (DCE) \cite{ong2020extreme}, and MR Fingerprinting (MRF) \cite{cao2022optimized}.

In parallel imaging \cite{pruessmann1999sense, griswold2002generalized, sodickson1997simultaneous}, a general approach to decrease the computation time is to reduce the number of coils actively used during reconstruction. One class of such methods is coil compression \cite{buehrer2007array,zhang2013coil,huang2008software}, where multi-channel data is linearly combined prior to reconstruction to obtain virtual coils sorted according to descending energy. Then, reconstruction is performed using only the significantly smaller subset of high-energy channels. In general, the number of high-energy virtual channels is a function of coil geometries and arrangement. However, in 3D Cartesian imaging, this number can be further decreased by leveraging the fully-sampled  k-space dimension (e.g. the readout direction) as in Geometric Coil Compression (GCC) \cite{zhang2013coil}. GCC leverages the fully-sampled dimension to spatially break down the 3D k-space into multiple 2D k-spaces through a Fourier transform along this spatial axis. Then, due to their spatial location, each 2D slice will have non-negligible signal in fewer coils, which allows compression to fewer virtual coils \cite{zhang2013coil}. However, this decomposition is not possible in 3D non-Cartesian imaging, where there is no fully-sampled axis. Thus, the 3D k-space data will have non-negligible signal in most of the coils, which hinders the ability of coil compression to reduce the number of channels.

Another line of work uses the stochastic gradient descent (SGD) method \cite{muckley2014accelerating,ong2020extreme}, which randomly selects subsets of coils at each iteration. While fully leveraging all data, this method requires constantly loading/unloading coil sensitivity maps to memory and often many iterations for convergence, which yields long reconstruction times \cite{bottou2012stochastic}. Furthermore, SGD is not versatile enough to be efficiently applied to the diversity of reconstruction formulations. 
For example, when the regularization function is non-smooth (e.g. $\ell_1$-norm) stochastic subgradient descent suffers from slow convergence \cite{bubeck2015convex}. Alternatively, proximal SGD (proxSGD), which has an improved convergence rate in this case \cite{bubeck2015convex}, is inefficient when the proximal operator of the regularization is computationally expensive \cite{kamilov2016parallel}.

Herein, we propose a novel, versatile, and computationally-efficient reconstruction method, referred to as coil sketching, which combines the best of both above mentioned approaches using a framework based on randomized sketching \cite{pilanci2015randomized,pilanci2016iterative,pilanci2017newton,tang2017gradient}. 
Randomized sketches are a group of well-established techniques for large-scale optimization with demonstrated success in machine learning and big data analysis \cite{pilanci2015randomized}. The basic idea uses a randomly-generated sketching matrix to project the problem onto a lower dimensional space, wherein the problem can be solved significantly faster and with a smaller memory footprint.

This article represents the first attempt to adapt randomized sketching theory to the MRI reconstruction problem. Coil sketching offers two main contributions:

\begin{enumerate}
    \item A novel, general, and versatile formulation of the reconstruction problem, using iterative sketching \cite{pilanci2016iterative}, that is readily applicable to nearly all MRI reconstruction problems.
    \item A structured sketching matrix design that leverages the Fourier structure of the MRI reconstruction problem and previous research on coil compression.
\end{enumerate}

\begin{figure*}[h]
\centerline{\includegraphics[width=2\columnwidth]{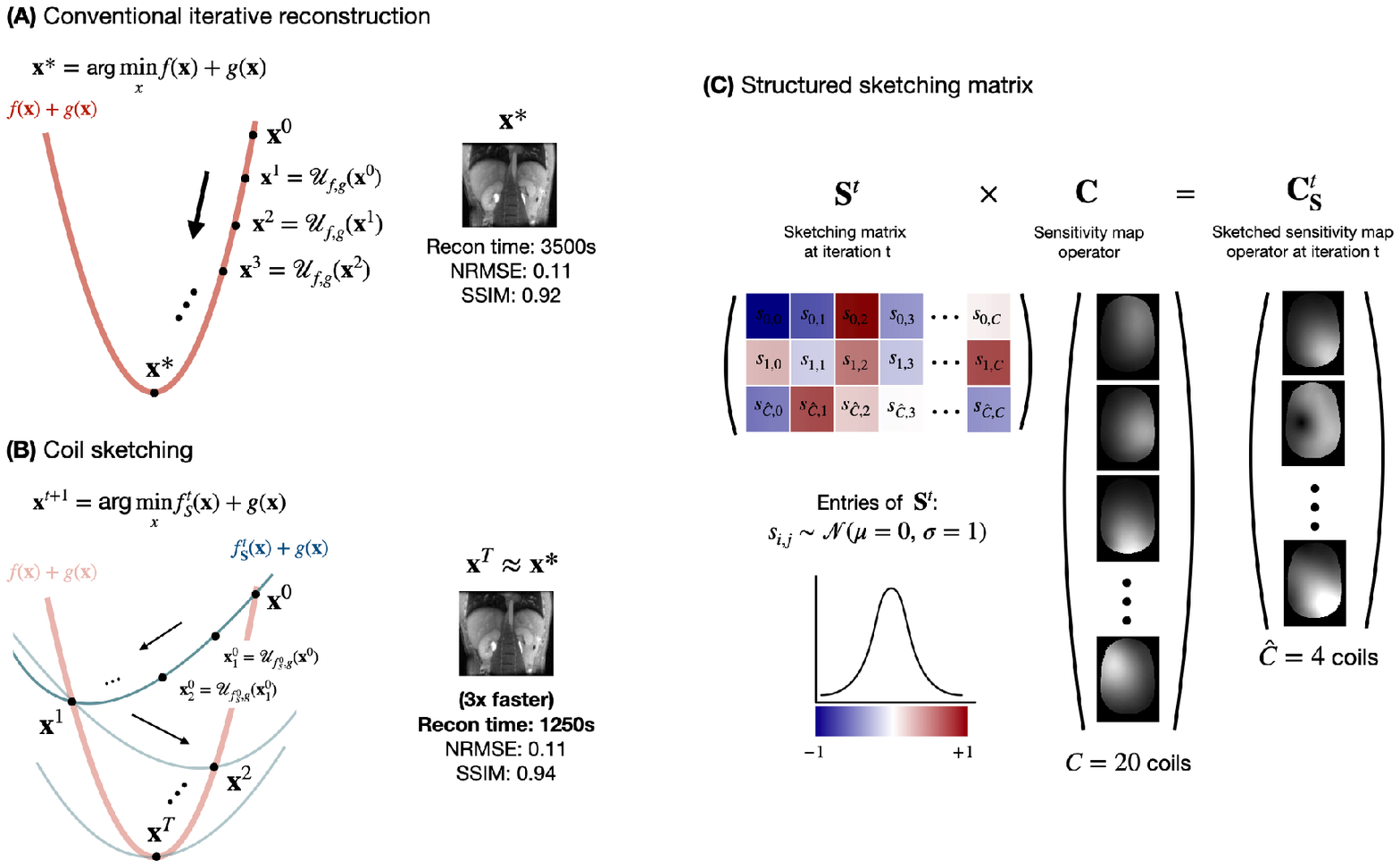}}
\caption[Coil sketching diagram]{\textbf{(A)} Conventional iterative image reconstruction recovers image $\mathbf{x}$ by directly minimizing an objective function $f(\mathbf{x}) + g(\mathbf{x})$, where $f(\mathbf{x})$ is the data consistency and $g(\mathbf{x})$ is a regularization function. Starting from an initial value $\mathbf{x^0}$, iterative algorithms yield a succession of approximate solutions ($\mathbf{x}^1$, $\mathbf{x}^2$, $\mathbf{x}^3$, ...), via an update rule $\mathcal{U}_{f,g}$, which converge to the solution $\mathbf{x^*}$.
\textbf{(B)} Coil sketching sequentially minimizes approximations $f^t_\mathbf{S}(\mathbf{x}) + g(\mathbf{x})$, where $f^t_\mathbf{S}(\mathbf{x})$ is a lower-dimensional approximation of the data consistency term. Similarly, the approximated problems can be solved iteratively via more computationally-efficient update rules $\mathcal{U}_{f^t_\mathbf{S},g}$. Each new solution $\mathbf{x}^t$ yields a more accurate reconstruction and eventually arrives at $\mathbf{x^T}\approx\mathbf{x^*}$. This approach enables higher computationally efficiency (3x faster in Results \ref{subsubsec:res_L1TV}, Figure \ref{fig6}) with virtually no penalty on reconstruction accuracy. \textbf{(C)} The approximations $f^t_\mathbf{S}(\mathbf{x})$ are formed by multiplying the problem with a randomly-generated structured sketching matrix $\mathbf{S}^t$ that produces random linear combinations only in the coil dimension and, effectively, reduces the number of coils concurrently used during reconstruction from $C$ to $\hat{C}$ ($C=20$ and $\hat{C}=4$ in this example).}
\label{fig1}
\end{figure*}

Figure \ref{fig1} illustrates both concepts. Early versions of this work were presented in References \cite{oscanoa2021coil, oscanoa2022coil}. This article is organized as follows. 
Section \ref{sec2} provides the relevant theory such as the MR image reconstruction problem, a brief introduction to randomized sketching, and the main concepts of our work, namely, the adaptation of sketching to MRI.
Then, Section \ref{sec3} elaborates on the experiments performed to validate performance and Section \ref{sec4} outlines the key results. Finally, Sections \ref{sec5} and \ref{sec6} provide discussion, conclusions, and future work.

\section{Theory}\label{sec2}

\subsection{Conventional reconstruction} \label{sec:conventional-recon}

We formalize the MRI reconstruction problem considered in this work. We consider the discrete multi-channel MRI forward model in Equation (\ref{eq:forward-model}), where the  image $\mathbf{x} \in \mathbb{C}^D$ is sampled at frequency points $\{f_i\}^N_{i=1}$ that yields $\emph{k}$-space measurements $\{ \mathbf{k}_c \in \mathbb{C}^N \}^C_{c=1}$ for each of the $C$ channels. Furthermore, we consider $C$-channel sensitivity maps $\{ \mathbf{c}_c \in \mathbb{C}^D \}^C_{c=1}$ and  white Gaussian noise vectors $\{ \mathbf{n}_c \in \mathbb{C}^N \}^C_{c=1}$.

\begin{equation}
\mathbf{k}_{c}[i]
=\frac{1}{\sqrt{D}} 
\sum_{d=0}^{D-1} \mathbf{c}_{c}[d] \mathbf{x}[d] \exp\left(-\imath 2 \pi f_{i} d / N \right)+\mathbf{n}_{c}[i]
\label{eq:forward-model}
\end{equation}

\noindent for $i \in \{ 1, ..., N \}$ and $c \in \{ 1, ..., C \}$.

We summarize the forward model as a linear model:

\begin{equation}
   \mathbf{k = FCx + n},
\end{equation}

\noindent where $ \mathbf{k} \in \mathbb{C}^{CN} $ and 
$\mathbf{n} \in \mathbb{C}^{CN}$ are stacked versions of 
$\{ \mathbf{k}_c \in \mathbb{C}^N \}^C_{c=1}$ and 
$\{ \mathbf{n}_c \in \mathbb{C}^N \}^C_{c=1}$ respectively, and the matrices $\mathbf{F}$ and $\mathbf{C}$ are defined as follows:

\begin{equation*}
\mathbf{C} = 
    \begin{bmatrix}
        \text{diag}\left( \mathbf{c}_1 \right) \\ 
         \vdots \\ 
        \text{diag}\left( \mathbf{c}_C \right)
    \end{bmatrix}, 
\end{equation*}

\begin{equation*}
    \mathbf{F} = \mathbf{I}_C \otimes \mathbf{\tilde{F}}, 
\end{equation*}

\noindent where $\otimes$ is the Kronecker product, $\mathbf{C} \in \mathbb{C}^{CD \times D}$ represents the point-wise multiplication by the $C$-channel sensitivity maps, and $\mathbf{F}\in \mathbb{C}^{CN\times CD}$ is a block diagonal matrix with $C$ blocks $\{ \tilde{\mathbf{F}} \in \mathbb{C}^{N \times D}\}$ that represent the Fourier transform.  In Cartesian imaging, $\tilde{\mathbf{F}} = \mathbf{U}\tilde{\mathbf{F}}_u$, where $\mathbf{U} \in \mathbb{R}^{N \times N}$ is an undersampling operator, and $\tilde{\mathbf{F}}_u \in \mathbb{C}^{N \times D}$ is the uniform Fourier transfom, implemented as the Fast Fourier Transform (FFT). In non-Cartesian imaging, $\tilde{\mathbf{F}} = \tilde{\mathbf{F}}_n \in \mathbb{C}^{N \times D}$, where $\tilde{\mathbf{F}}_n$ is the non-uniform Fourier transform, implemented as the Non-Uniform FFT (NUFFT) \cite{beatty2005rapid}. Additionally, $\mathbf{I}_C \in \mathbb{R}^{C \times C}$ is the identity matrix. We will use the notation $\mathbf{I}_a$, where $a \in \mathbf{R}$, to refer to the $a \times a$ identity matrix.
\subsubsection{Problem Formulation}\label{sec:conventional-problem}
Our method is applicable to both Cartesian and non-Cartesian reconstruction problems, which we formalize next.
Given the acquired \emph{k}-space measurements $\mathbf{k}$, image reconstruction aims to recover the unaliased image by solving a regularized least squares optimization problem:

\begin{equation}
\label{eq:conventional_recon}
    \mathbf{x}^* = \argmin_{\mathbf{x}} f(\mathbf{x})+g(\mathbf{x})
\end{equation}

\begin{equation}
\label{eq:conventional_recon_dc}
    f(\mathbf{x}) = \frac{1}{2}\|\mathbf{A} \mathbf{x}-\mathbf{y}\|_{2}^{2} 
\end{equation}

\noindent where $f(\mathbf{x})$ is the data consistency term, $g(\mathbf{x})$ is the regularization function. In Cartesian imaging, $\mathbf{A}=\mathbf{FC}$ and $\mathbf{y}=\mathbf{k}$. In non-Cartesian imaging, we consider reconstruction with density compensation \cite{pipe1999sampling}:
\begin{equation*}
\mathbf{A} = \mathbf{W}^{1/2}\mathbf{FC},
\end{equation*}
\begin{equation*}
\mathbf{y} = \mathbf{W}^{1/2}  \mathbf{k}, 
\end{equation*}
\noindent where $\mathbf{W} \in \mathbb{R}^{CN}$ is a diagonal matrix that applies gridding density compensation, which has been demonstrated to improve condition number of reconstruction for gradient-based methods \cite{baron2018rapid}. Since $\mathbf{W}$ applies identical density compensation to each coil, we can also define:
\begin{equation*}
\mathbf{W}^{1/2} = \mathbf{I}_C \otimes\text{diag}(\mathbf{w})^{1/2}
\end{equation*}
\noindent where $\mathbf{w} \in \mathbb{R}^N$ is the density compensation function.

In particular, we will consider three types of regularization for our experiments:

\begin{enumerate}
    \item $\ell_2$-Norm \cite{ying2004tikhonov}: $g(\mathbf{x}) = \frac{\lambda}{2} \|\mathbf{x}\|_{2}^{2}$
    \item $\ell_1$-Wavelets \cite{lustig2007sparse}: $g(\mathbf{x}) = \lambda \| \mathbf{\Psi} \mathbf{x}\|_{1}$, where $\mathbf{\Psi} \in \mathbb{C}^{D \times D}$ is a unitary wavelet transform operator.
    \item $\ell_1$-Total variation ($\ell$1-TV) \cite{block2007tv1,ma2008tv2} : $g(\mathbf{x}) = \lambda \| \mathbf{T} \mathbf{x}\|_{1}$, where $\mathbf{T} \in \mathbb{C}^{D \times D}$ is a first-order finite difference operator with periodic boundary extension.
\end{enumerate}

Respectively, the resulting optimization problems can be solved efficiently using:

\begin{enumerate}
    \item Conjugate Gradient (CG) \cite{shewchuk1994introduction, pruessmann2001advances,maier2021cgsense} 
    \item Fast Iterative Soft-Thresholding Algorithm (FISTA) \cite{beck2009fast}
    \item Primal Dual Hybrid Gradient (PDHG) \cite{chambolle2011first}
\end{enumerate}

Although these algorithms have considerable differences, in the context of MR image reconstruction, the largest computational cost of each algorithm is the application of matrix $\mathbf{A}$.  In practice, matrix $\mathbf{A}$ is never explicitly formed. Instead, we leverage the matrix structure to reduce the computational complexity. For example, in non-Cartesian imaging we apply the following coil-wise operation:
\begin{equation}
    \mathbf{y}_c = A_c(\mathbf{x}) = \mathbf{w}^{\odot 1/2} \odot \mathcal{F}\{ \mathbf{c}_c  \odot \mathbf{x} \},
\end{equation}
\noindent where $\odot$ is the Hadamard product, $\mathcal{F}$ is the NUFFT, and $\mathbf{w}^{\odot 1/2}$ is the Hadamaard squared root of vector $\mathbf{w}$. Consequently, the computational cost of the forward model increases with the number of coils, $C$. 
%Use hadamaard square root
\subsection{Randomized sketching} \label{sec:sketching} 

% Similarly to the MRI reconstruction problem (Section \ref{sec:conventional-recon}), randomized sketches consider a constrained least squares problem:
Similarly to MRI reconstruction, randomized sketching considers a general constrained least squares problem:

\begin{equation}
\label{eq:original_lsq}
    \mathbf{u}^* = \argmin_{\mathbf{u} \in \mathcal{C}}  \|\mathbf{B u} -\mathbf{v}\|_{2}^{2},
\end{equation}

\noindent where $\mathbf{u}^* \in \mathbb{R}^D$ is the solution, $\mathbf{v} \in \mathbb{R}^N$ is a data vector, $\mathbf{B} \in \mathbb{R}^{N \times D}$ is a data matrix and $\mathcal{C}$ is a convex set. In the most basic case, $\mathcal{C}$ would be $\mathbb{C}^D$, but literature considers many other types such as $\ell_1$-balls, and $\ell_2$-balls, which can be converted to the regularizations in Subsection \ref{sec:conventional-recon} via the Lagrangian.

\subsubsection{Classical Sketching}
The classical sketch \cite{pilanci2015randomized} obtains an approximate solution $\mathbf{\tilde{u}}$ to equation (1). Sketching uses a randomly generated matrix $\mathbf{S} \in \mathbb{R}^{\hat{N} \times N}$ ($\hat{N} \ll N$) to project the problem onto a lower dimensional space. Herein, we consider two of the most common sketching matrices $\mathbf{S}$: (1) independent and identically distributed (i.i.d) standard Gaussian entries and (2) i.i.d. Rademacher entries (uniformly distributed over $\{ -1,+1 \} $). 

The new sketched least squares problem is:

\begin{equation}
    \mathbf{\tilde{u}} = \argmin_{\mathbf{u} \in \mathcal{C}}\|\mathbf{SBu} -\mathbf{Sv}\|_{2}^{2},
\end{equation}
where $\mathbf{B}$ and $\mathbf{v}$ are approximated by their sketched versions $\mathbf{SB} \in \mathbb{R}^{\hat{N} \times D}$ and $\mathbf{Sv} \in \mathbb{R}^{\hat{N}}$. This new approximated problem can be solved with substantially less computational cost at the expense of reduced accuracy. In the context of medical imaging, reconstruction inaccuracies could lead to considerable image quality loss and substantial artifacts. Therefore, we also consider more advanced sketching techniques.

\subsubsection{Iterative Hessian Sketching}
The Iterative Hessian Sketch \cite{pilanci2016iterative} (IHS) sequentially approximates the problem multiple times around the current estimate $\mathbf{u}^t$. Each subsequently approximated problem will yield a new estimate $\mathbf{u}^{t+1}$ with higher accuracy. The new optimization problem is:

\begin{equation}
\begin{split}
\label{eq:hessian_sketch}
     \mathbf{u}^{t+1} = \argmin_{\mathbf{u} \in \mathcal{C}}
    \frac{1}{2}\|\mathbf{S}^{t}&\mathbf{B}(\mathbf{u}-\mathbf{u}^t )\|_{2}^{2} + \langle \mathbf{u}, \mathbf{B}^\mathsf{H}(\mathbf{B}\mathbf{u}^t -\mathbf{v}) \rangle,
\end{split}
\end{equation}
where $\langle \cdot,\cdot \rangle$ is the inner product, and $\mathbf{S}^{t} \in \mathbb{R}^{\hat{N}\times N}$ is the randomly-generated sketching matrix used at iteration $t$. The objective function in equation (\ref{eq:hessian_sketch}) can be interpreted as a second-order Taylor approximation of the original objective function in equation (\ref{eq:original_lsq}) around the current estimate $\mathbf{u}^t$ with a sketched Hessian $(\mathbf{SB})^\mathsf{T}(\mathbf{SB})$.  Section S1 of Supporting Material provides a detailed derivation. This optimization problem is slightly more computationally costly to solve, but yields significantly more accurate results since, according to theory, the estimates $\mathbf{u}^{t+1}$ will converge to the original  $\mathbf{u}^*$ geometrically with the iteration number $t$ \cite{pilanci2016iterative}.

\subsection{Sketched reconstruction} \label{sec:sketched-recon}

We propose the use of randomized sketching to reduce the computational cost of image reconstruction. Our work proposes two main concepts. First, we reformulate the reconstruction problem using iterative sketching \cite{pilanci2016iterative,tang2017gradient}. Second, we propose a structured sketching matrix design that synergistically leverages the computational efficiencies of sketching, the Fourier structure of the problem, and previous research on coil compression.

\subsubsection{Sketched Problem Formulation}
Instead of solving the reconstruction problem directly, we break the original problem, equations (\ref{eq:conventional_recon}, \ref{eq:conventional_recon_dc}), into multiple sub-problems. The sketched sub-problem at iteration $t$ is:

\begin{equation}
\label{eq:sketched_recon}
    \mathbf{x}^{t+1} = \argmin_{\mathbf{x}} f^t_\mathbf{S}(\mathbf{x})+g(\mathbf{x}),
\end{equation}

\noindent where $f^t_{\mathbf{S}}$ is the sketched data consistency term:

\begin{equation}
\label{eq:sketched_recon_dc}
    f^t_\mathbf{S}(\mathbf{x}) = 
    \frac{1}{2}\|\mathbf{A}^t_\mathbf{S} (\mathbf{x}-\mathbf{x}^t )\|_{2}^{2} + \langle \mathbf{x}, \mathbf{A}^\mathsf{H}(\mathbf{A}\mathbf{x}^t -\mathbf{y}) \rangle,
\end{equation}

\noindent where $\mathbf{A}^t_{\mathbf{S}} = \mathbf{S}^t \mathbf{A} \in \mathbb{C}^{\hat{C}N \times D}$ ($\hat{C} < C$) is the sketched forward model at iteration $t$. $\mathbf{A}^t_{\mathbf{S}}$ is a lower-dimensional approximation of the original forward model $\mathbf{A}$ and its calculation is detailed in the next Subsection \ref{sub:structured-sketching-matrix}.

% The final reconstructed image is the last iterate $\mathbf{x}^* \approx \mathbf{x}^T$.
As a great advantage, the new sketched reconstruction problem, equations (\ref{eq:sketched_recon}, \ref{eq:sketched_recon_dc}), allows the use of the same solver as the original reconstruction problem, equations (\ref{eq:conventional_recon}, \ref{eq:conventional_recon_dc}), since it only modifies the data consistency term from a least squares term to a similar quadratic term. This trait makes the method highly versatile, and can be readily applied to nearly all reconstruction problems with the form of a regularized or constrained least squares formulation.%, e.g. low-rank, $\ell_1$, $\ell_2$ methods, etc.

\subsubsection{Structured Sketching Matrix}\label{sub:structured-sketching-matrix}
A conventional sketching matrix, $\mathbf{S}$, would break the Fourier structure of the forward model, $\mathbf{A}$, preventing us from leveraging the high computational efficiency of the FFT. Thus, we propose a structured sketching matrix design, $\mathbf{S}^t$, which utilizes sketching to reduce computational burden while also preserving the Fourier structure of the problem. The key idea is that $\mathbf{S}^t$ will produce random linear combinations only in the coil dimension, reducing the effective number of coils actively used during reconstruction. Figure 1 depicts this procedure.

We define the structured matrix, $\mathbf{S}^t \in \mathbb{C}^{\hat{C}N \times CN}$ ($\hat{C} < C$), as:

\begin{equation}
    \mathbf{S}^t = \mathbf{\tilde{S}}^t \otimes \mathbf{I}_N,
\end{equation}

\noindent where $\mathbf{\tilde{S}}^t \in \mathbb{R}^{\hat{C} \times C}$ is a random matrix with i.i.d. entries , $\mathbf{I}_N \in \mathbb{R}^{N \times N}$ is the identity matrix, and $\otimes$ is the Kronecker product. When applied to model matrix $\mathbf{A}$, effectively, this structured matrix will produce random linear combinations in the coil dimension. Therefore, since $\mathbf{W}$ and $\mathbf{F}$ are coil-wise operations, the order of the operators can be exchanged as follows:

\begin{equation}
\begin{split}
    \mathbf{A}^t_\mathbf{S} = \mathbf{S}^t\mathbf{W}^{1/2}\mathbf{FC}  
                 = \mathbf{\hat{W}}^{1/2} \mathbf{\hat{F}} \mathbf{S}^t \mathbf{C}  
                = \mathbf{\hat{W}}^{1/2} \mathbf{\hat{F}}\mathbf{C}^t_\mathbf{S},
\end{split}
\end{equation}

\noindent where $\mathbf{\hat{W}} \in \mathbb{C}^{\hat{C}M \times \hat{C}M}$, $\mathbf{\hat{F}} \in \mathbb{C}^{\hat{C}M \times \hat{C}M}$ are similar to $\mathbf{W}$ and $\mathbf{F}$ but with reduced number of dimensions in the coil dimension ($\hat{C} < C$). $\mathbf{C}^t_\mathbf{S}$ is a sketched sensitivity map operator composed by a new reduced set of sensitivity maps $\{ \mathbf{\hat{c}}_c \in \mathbb{C}^D \}^{\hat{C}}_{c=1}$ :

\begin{equation}
    \begin{bmatrix}
         \mathbf{\hat{c}}_1[d] \\ 
         \vdots \\ 
         \mathbf{\hat{c}}_{\hat{C}}[d]
    \end{bmatrix}
    = \mathbf{\tilde{S}}^t \times 
    \begin{bmatrix}
         \mathbf{c}_1[d] \\ 
         \vdots \\ 
         \mathbf{c}_C[d]
    \end{bmatrix}
    \text{, for } d = 1, ..., D.
\end{equation}

%Supplemental Material 1 shows a more detailed demonstration. 
The computational cost of the new sketched forward model, $\mathbf{A}^t_\mathbf{S}$, will be reduced by the reduced number of coils considered, $\hat{C}$. Finally, we acknowledge that our proposed structured sketching matrix imposes dependencies on the randomly-generated sketching matrix, which prevents us from maintaining the formal theoretical guarantees studied in other works \cite{pilanci2015randomized,pilanci2016iterative}. However, this structure is essential for computational speed since it allows to keep leveraging the high computational efficiency of the FFTs. We leave the analysis of this matrix design for future work.
%to $\mathcal{O}(\hat{C}N \log{N})$, i.e. proportional to the reduced number of coils $\hat{C}$.

\subsubsection{Inclusion of Virtual Coils}
Random projection onto a lower-dimensional space reduces the computational cost of the optimization problem at the expense of reduced accuracy. Analogously, randomly combining the coils into fewer sketched coils will reduce computational cost of the forward model at the expense of  accuracy/information, since the sketched coils will be an approximation of the original coils. On the other hand, coil compression works have shown that a considerable amount of low-energy virtual coils can be discarded without significant loss of image quality \cite{buehrer2007array,zhang2013coil,huang2008software}. Therefore, we hypothesize that a sketching matrix design that only loses information in the low-energy virtual coils would be a better approximation than a design that loses information in all virtual coils. 

We define $V \in \mathbb{R}$ and $S\in \mathbb{R}$ as the number of high-energy virtual coils and the number of sketched low-energy coils, respectively, included in the sketched coil sensitivity map operator, $\mathbf{C}^t_\mathbf{S}$, where $\hat{C} = V + S$. Now, $\tilde{\mathbf{S}}^t$ has the following form:

\begin{equation*}
\tilde{\mathbf{S}}^t = 
    \begin{bmatrix}
        \mathbf{I}_{V} & \mathbf{0} \\ 
        \mathbf{0} & \mathbf{\tilde{S}}_{S}^t
    \end{bmatrix},
\end{equation*}

\noindent where $\mathbf{\tilde{S}}_{S}^t \in \mathbb{R}^{S \times (C-V)}$ is a random matrix with i.i.d. entries, and $\mathbf{0}$ are zero matrices. 
We test our hypothesis empirically through the experiments in Subsection 3.2. 

\subsubsection{Algorithm}
Coil Sketching is detailed in Algorithm \ref{alg1}. First, we apply coil compression and keep the first $\hat{C}_0$ high-energy coils. This step will reduce the computational cost of the true gradient calculation (line 10) and the sketched model formation (line 11). We chose $\hat{C}_0$ to avoid significant loss of image quality. Next, we include an optional initialization step similar to other iterative sketching algorithms, where we run a fast, yet inaccurate, reconstruction using the classical sketch formulation. 

Then, we proceed with the iterative sketch-based reconstruction. We estimate the true gradient $\mathbf{d}$ using the current estimate $\mathbf{x}^t$ and we form the sketched model $\mathbf{A}_\mathbf{S}^t$ which contains only $\hat{C}$ coils. We solve the sketched reconstruction problem using the same optimizer as the original problem. Implementation details can be found in Section S2 of the Supporting Material. The cost of the new sketched forward operator, $\mathbf{A}^t_\mathbf{S}$, defines the computational cost of coil sketching (\emph{i.e.}, the computational cost reduces with the new reduced number of coils, $\hat{C}$, in $\mathbf{C}^t_\mathbf{S}$).

\begin{algorithm}[h]
\caption{Coil sketching}\label{alg1}
\begin{algorithmic}[1]
    \State \textbf{Input:} \emph{k}-space measurements $\mathbf{y}$
    \State \textbf{Output:} Reconstructed image $\mathbf{x}^* \approx \mathbf{x}^T$
\State Define parameters $\hat{C}_0$ and $\hat{C}$ 
\State Perform coil compression and keep the first $\hat{C}_0$ coils 
\State Set initial estimate: $\mathbf{x}^0$, \emph{e.g.} $\mathbf{x=0}$ 

\If{Initialization}
    \State $\mathbf{x}^{1} = \argmin_\mathbf{x} \frac{1}{2}\|\mathbf{S}^0 \mathbf{A}\mathbf{x}-\mathbf{S}^0\mathbf{y} \|_{2}^{2} + g(\mathbf{x}) $
\EndIf
\For{$t = 1,...,T$}
    \State Compute true gradient: $\mathbf{d} = \mathbf{A}^\mathsf{H}(\mathbf{A}\mathbf{x}^t -\mathbf{y})$
    \State Form sketched model with $\hat{C}$ coils:  $\mathbf{A}^{t}_\mathbf{S} = \mathbf{S}^{t} \mathbf{A}$
    \State Solve lower-dimensional sketched problem:
    \State $\;\; \mathbf{x}^{t+1} =  \argmin_{\mathbf{x}} \frac{1}{2}\| \mathbf{A}^{t}_\mathbf{S}(\mathbf{x}-\mathbf{x}^t )\|_{2}^{2} +  \langle \mathbf{x}, \mathbf{d} \rangle + g(\mathbf{x})$
\EndFor
\end{algorithmic}
\end{algorithm}

\section{Methods}\label{sec3}

Our experiments are divided into two main groups as shown in Table \ref{table1}. The first group evaluates different design options ($V$, $S$, and random distributions) for the sketching matrix $\mathbf{\tilde{S}}^t$ based on image quality metrics and the inverse g-factor. We use the relatively smaller 2D datasets for computational simplicity.  We perform two experiments. The first experiment evaluates design choices through the objective function value and image quality metrics. We consider the 2D radial dataset and the most common regularization in PICS reconstruction: $\ell_1$-Wavelets. The second experiment evaluates through the inverse g-factor, a common approach to measure SNR performance in parallel imaging. Since g-factor analysis can only be performed in linear reconstructions, we consider the 2D Cartesian dataset with regular undersampling and reconstruction with $\ell 2$-norm regularization.

With the selected sketching matrix design, the second group of experiments evaluates performance of Coil Sketching in our high-dimensional application of interest: accelerated 3D Cones reconstruction \cite{gurney2006design}. We evaluate the two main advantages of our method: increased computational efficiency and versatility. We measure computational efficiency through convergence time, memory footprint, and image quality metrics. Whereas we demonstrate versatility by benchmarking Coil Sketching in two of the most common PICS reconstructions: $\ell_1$-Wavelets and $\ell_1$-Total Variation.
Different regularization functions have different solver choices for efficient reconstruction. Coil sketching allows re-using the same solver as the original reconstruction problem, which makes it readily applicable to virtually any reconstruction problem. Conversely, other methods such as SGD might not be efficiently applicable to all reconstruction problems.

All the code for the experiments was implemented in Python using the Sigpy library \cite{ong2019sigpy}. The reconstructions were run in GPU memory (NVIDIA 24GB Titan RTX).

\subsection{Dataset Details}
All imaging was performed with Institutional Review Board (IRB) approval and consent.
To develop and evaluate the sketching algorithm we used five datasets: 
\begin{itemize}
    \item One 3D Cartesian knee acquisition
\end{itemize}

\begin{landscape}
\begin{table}
\centering
\includegraphics[width=\linewidth]{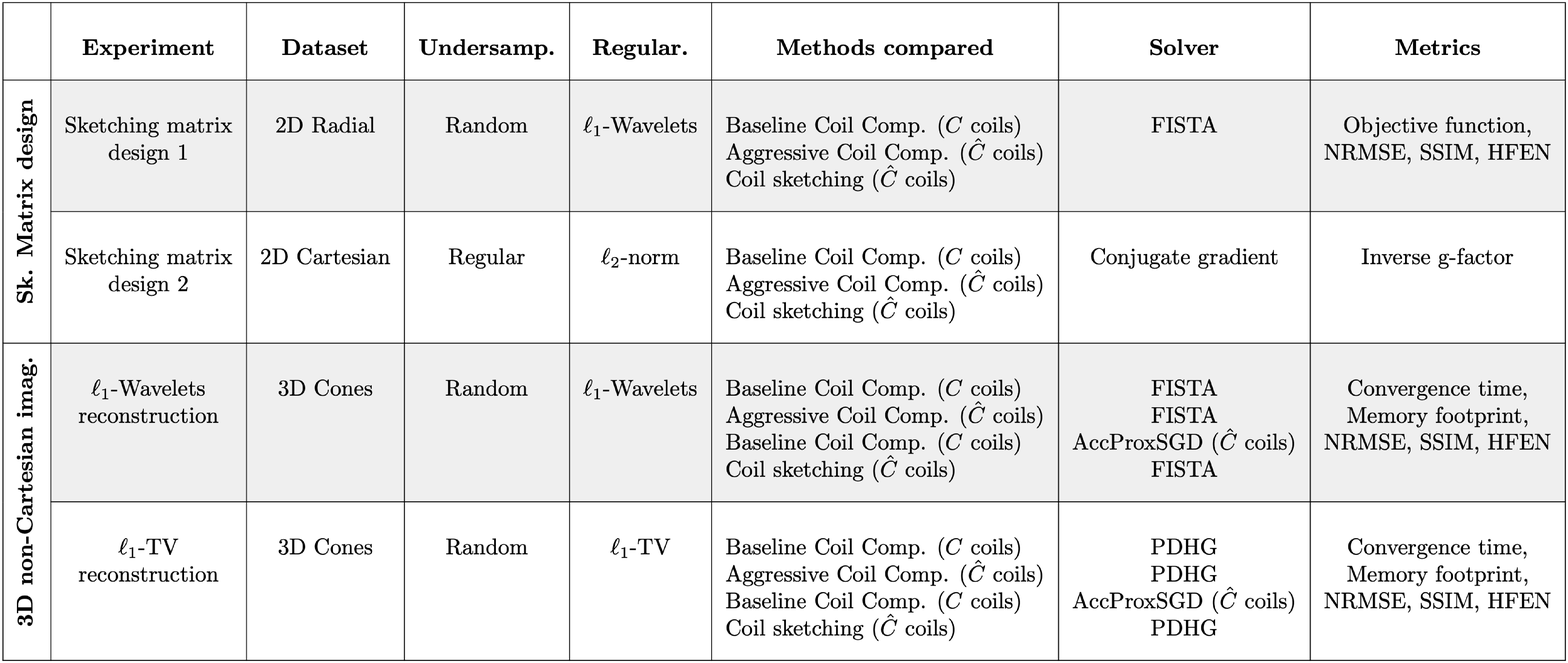}
\caption[Summary of experiments performed]{
Summary of experiments performed.  We have two main group of experiments. The first group (Subsection 3.2) evaluates different design choices for the structured sketching matrix using relatively smaller 2D datasets. We measure objective function value, image quality metrics, and inverse g-factor. The second group (Subsection 3.3) benchmarks Coil sketching in our high-dimensional application of interest: 3D Cones reconstruction. We measure computational efficiency through convergence time and memory footprint, and image quality metrics.}
\label{table1}
\end{table}
\end{landscape}

\begin{itemize}
    \item One 2D radial liver acquisition
    \item Two 3D cones abdominal acquisitions
\end{itemize}

The fully-sampled 3D Cartesian fast-spin echo multi-coil knee MRI dataset is publicly available on mridata.org \cite{ong2018mridata}. 
The dataset had eight virtual channels and image size $320 \times 320 \times 256$. Each 2D $320 \times 256$ slice was treated as a separate example.

The radial liver dataset was acquired on a 3.0 T scanner (GE HealthCare, Waukesha, WI) using a 32-channel abdominal array. The FOV is $370 \times 370\; \text{mm}^2$ with a receiver bandwidth of $\pm$250 kHz and TR/TE=3.4/1.5 ms. The trajectory consisted of 402 radial spokes with tiny golden angle increments and 512 readout points per spoke. A total of 10 slices were acquired where each slice was treated as a separate example. The reconstructed image size is $256 \times 256$. 

The first 3D cones \cite{gurney2006design} abdominal dataset was acquired on a 1.5 T scanner (GE HealthCare, Waukesha, WI) using a 40-channel torso array. The FOV is $240 \times 240 \times 220 \; \text{mm}^3$ with a receiver bandwith of $\pm$125 kHz and TR/TE-40.1/0.11 ms. The trajectory consisted of 36,163 readouts with 615 readout points each. The reconstructed image size is $360\times256\times197$.

The second 3D cones \cite{gurney2006design} abdominal dataset was acquired on a 1.5 T scanner (GE HealthCare, Waukesha, WI) using a 35-channel array torso coil. The FOV is $300 \times 300 \times 240 \; \text{mm}^3$ with receiver bandwidth of $\pm$125 kHz and TR/TE=26.9/0.03 ms. The trajectory consists of 50,377 readouts with 513 readout points each. The reconstructed image size is $425\times242\times264$. 

For computational simplicity, all multi-channel datasets underwent an initial SVD-based coil compression and the lower-energy channels were discarded. 
 
For all datasets, the multi-channel \emph{k}-space data underwent an initial SVD-based coil compression and the lower-energy virtual channels were discarded to reduce computational cost. For notation simplicity, in the remaining of this work, we will denote $C$ as the number of virtual channels kept after this initial coil compression. For the 2D cases, we considered the first $C=8$ virtual channels, which allowed us to maintain $95\%$ of the total energy. For the 3D cases, we considered the first $C=20$ virtual channels, which allowed us to maintain $98\%$ of the total energy. The coil sensitivity maps were estimated using ESPIRiT \cite{uecker2014espirit} with automatic parameter selection \cite{iyer2020sure}.

\subsection{Sketching matrix design}
We evaluate two main parameters: 
% in the design of the sketching matrix $\mathbf{\tilde{S}}^t$:

\begin{enumerate}
    \item Probability density function for the random entries of $\mathbf{\tilde{S}}_{S}^t$, we consider two distributions used in the literature: Gaussian ($\mathcal{N}(\mu=0,\sigma=1/S)$) and Rademacher ($\{ +1, -1\}$ entries with equal probability $p=0.5$).
    \item Value of $V$, we vary $V$ from 0 to $\hat{C}$.
\end{enumerate}

Additionally, we compare coil sketching to a baseline reconstruction using all $C=8$ coils and aggressive coil compression using a reduced number of coils $\hat{C}$. $\hat{C}$ ranges from two to six. We perform two main experiments:

\subsubsection{2D $\ell_1$-Wavelets Reconstruction}
We retrospectively undersample ($R=2$, random undersampling) the radial dataset and perform $\ell_1$-Wavelets reconstruction on each slice separately. For each design option, we run 200 Coil Sketching with $\hat{C}$ coils. Each run has a different random seed initialization. 

We evaluate performance quantitatively through the objective function value and image quality metrics. We measure the Normalized Root Mean Squared Error (NRMSE), the Structural Similarity Index (SSIM) \cite{wang2004image}, and the High Frequency Error Norm (HFEN) \cite{langkammer2018qsm} with respect to the fully-sampled data. Additionally, we evaluate the resulting images qualitatively by visual inspection.

\subsubsection{2D $\ell_2$ Inverse g-factor Analysis}
% One of the main drawbacks of aggressive coil compression is the loss of energy, which produces shading artifacts and, also, SNR loss in parallel imaging reconstruction. This noise amplification can be measured by the inverse g-factor. 
We assess SNR impact using the inverse g-factor\cite{pruessmann1999sense}.
We use a Monte Carlo based method \cite{robson2008comprehensive} with 60 trials. At each trial, the retrospectively undersampled ($R=3$, regular) Cartesian dataset is injected with random Gaussian noise and reconstructed using $\ell_2$-norm reconstruction on each slice separately. We measure the inverse g-factor of Coil Sketching with a reduced number of coils, $\hat{C}$, for each design option.

% \begin{table*}[h]
% \caption{Summary of experiments performed}
% \label{table1}
% % \setlength{\tabcolsep}{}
% \begin{tabular}{|p{75pt}|p{60pt}|c|c|p{60pt}|p{85pt}|}
% \hline
% \textbf{Experiment} & \textbf{Dataset} & \textbf{Undersampling} & \textbf{Regularization} &\textbf{Solver} & \textbf{Metrics}\\
% \hline
% \rowcolor{gray!25}Sketching matrix design 1 & 2D radial & random & $\ell_1$-Wavelets & FISTA & Objective function, NRMSE, SSIM\\
% Sketching matrix design 2 & 2D cartesian & regular & $\ell_2$ & CG & Inverse g-factor \\
% \rowcolor{gray!25}$\ell_1$-Wavelets reconstruction & 3D cones & random & $\ell_1$-Wavelets & FISTA, AccProxSGD & Convergence time, NRMSE, SSIM\\
% $\ell_1$-TV reconstruction & 3D cones & random & $\ell_1$-TV & PDHG, AccProxSGD & Convergence time, NRMSE, SSIM\\
% \hline
% \end{tabular}
% \label{tab1}
% \end{table*}

% \usepackage{multirow}

% \usepackage{multirow}

% \usepackage{multirow}

\subsection{3D non-Cartesian imaging}
\subsubsection{3D $\ell_1$-Wavelets Reconstruction}

With the best design chosen, we evaluate the performance of Coil Sketching in our application of interest: accelerated 3D cones reconstruction \cite{gurney2006design}. This case study is of main interest due to the massive consumption of computational resources in comparison to 2D datasets.

We assess one of the most common Compressed Sensing reconstructions, $\ell_1$-Wavelets reconstruction.
We retrospectively undersample the 3D cones datasets ($R=2.5$) and reconstruct them with $\ell_1$-Wavelets Daubechies 4 regularization. We consider reconstructions with all $C=20$ coils and reduced number of coils, $\hat{C}$, ranging from two to ten. We evaluate the following methods:
\begin{itemize}
    \item Baseline reconstruction with all $C$ virtual coils 
    \item Aggressive coil compression with $\hat{C}$ virtual coils 
    \item Accelerated Proximal Stochastic Gradient Descent (AccProxSGD) with coil batch size $\hat{C}$
    \item Coil Sketching with reduced number of coils $\hat{C}$
\end{itemize}

The baseline reconstruction used $C$ virtual channels. Due to memory constraints, we performed coil batching, i.e. the reconstruction sequentially loaded in memory $\hat{C}$ coils at a time during reconstruction. This methodology slowed the reconstruction process; however, it was necessary in order to perform the reconstruction in GPU memory. We used the FISTA \cite{beck2009fast} solver and $\lambda = 0.02$, empirically tuned using a grid search \cite{alibrahim2021hyperparameter,lustig2007sparse}.

The aggressive coil compression method considered a reduced number of coils, $\hat{C}$, and used the FISTA solver \cite{beck2009fast}. To account for the significant energy loss in the data consistency term, we re-tuned $\lambda$ to $\lambda= 0.01$. Furthermore, we implemented AccProxSGD as previously described \cite{muckley2014accelerating,ong2020extreme}, including the proximal step for $\ell_1$-Wavelets regularization and Nesterov Momentum.
We used the decreasing step size rule: $\alpha^t = \max( \beta^t \times \alpha_0, \beta_{min} )$ and the original $\lambda = 0.02$. We ran 20 reconstructions and report the best performance for comparison.

Finally, we performed Coil Sketching reconstruction with a reduced number of coils, $\hat{C}$. We used the FISTA solver \cite{beck2009fast} for the sketched subproblems, and $\lambda = 0.02$. For computational speed, we used the same step size for all the subproblems, we estimated the step size from the first subproblem. Empirically, we found that this heuristic yielded equivalent convergence rates.

We tested values of $\hat{C}$ ranging from four to ten. We evaluate performance through convergence speed, peak memory usage, and reconstructed image quality. Convergence speed is measured by the normalized distance to $x^\infty$: $ d(\mathbf{x}^t) = \|\mathbf{x}^t - \mathbf{x}^{\infty}\|_{2}/\|\mathbf{x}^{\infty}\|_{2}$. We consider the algorithm converged when within 5\% distance. Furthermore, we monitored GPU memory footprint through CuPy's built-in libraries and the nvidia-smi command. Finally, we evaluate image quality versus fully-sampled reference quantitatively through NRMSE, SSIM \cite{wang2004image}, and HFEN \cite{langkammer2018qsm} versus the fully-sampled images, and qualitatively through visual inspection.

\subsubsection{3D $\ell_1$-TV Reconstruction }
One main advantage of Coil Sketching is the versatility to efficiently adapt to virtually any type of reconstruction. We also assess the method for $\ell_1$-TV reconstruction. We retrospectively undersample the 3D cones datasets (R=2.5) and reconstructed them with $\ell_1$-TV regularization. We consider the same values of $C$ and $\hat{C}$ and the same four reconstruction methods. 

Similarly to the $\ell_1$-Wavelets experiment, the baseline reconstruction used $C$ virtual channels  and coil batching of size $\hat{C}$. We used the Primal Dual Hybrid Gradient (PDHG) \cite{chambolle2011first} as the solver with $\lambda=0.01$, tuned empirically by grid search \cite{alibrahim2021hyperparameter,lustig2007sparse}. Likewise, aggressive coil compression considered a reduced number of coils $\hat{C}$, used PDHG as solver, and re-tuned $\lambda=0.005$. 
AccProxSGD considered $\hat{C}$ coils, $\lambda=0.01$, and solved the proximal step with PDHG \cite{chambolle2011first}.
Coil Sketching considered $\hat{C}$ coils, used the PDHG as solver, and $\lambda=0.01$. We evaluated performance similarly to the $\ell_1$-Wavelets case. 

\section{Results}\label{sec4}

\subsection{Sketching Matrix Design}

Figure 2 shows the results of the first sketching matrix design experiment for a reduced number of coils $\hat{C}=3$ (Figure 2A) and $\hat{C}=4$ (Figure 2B). We chose a low number of coils $\hat{C}$ to demonstrate the ability of coil sketching to significantly reduce the number of coils and, thus the computational cost of reconstruction, without compromising reconstruction accuracy. We normalized the objective function and NRMSE values using the values of the baseline reconstruction with $C=8$ virtual coils. Additionally, Supporting Figure S1 shows similar results for the normalized SSIM and HFEN metrics, and Supporting Figure S2 shows results for a high number of coils $\hat{C}=6$.

The Rademacher distribution presents less variance and more precise reconstructions when compared to the Gaussian distribution. Furthermore, for both distributions, the precision and accuracy are increased as we increment the value of $V$ (number of high-energy virtual coils included). However, when $V=\hat{C}$ ($S=0$), the optimization diverges, which suggests that, in this case, the sketched subproblems are not accurate approximations of the original problem.

\begin{figure*}[h]
\centerline{\includegraphics[width=1.37\columnwidth]{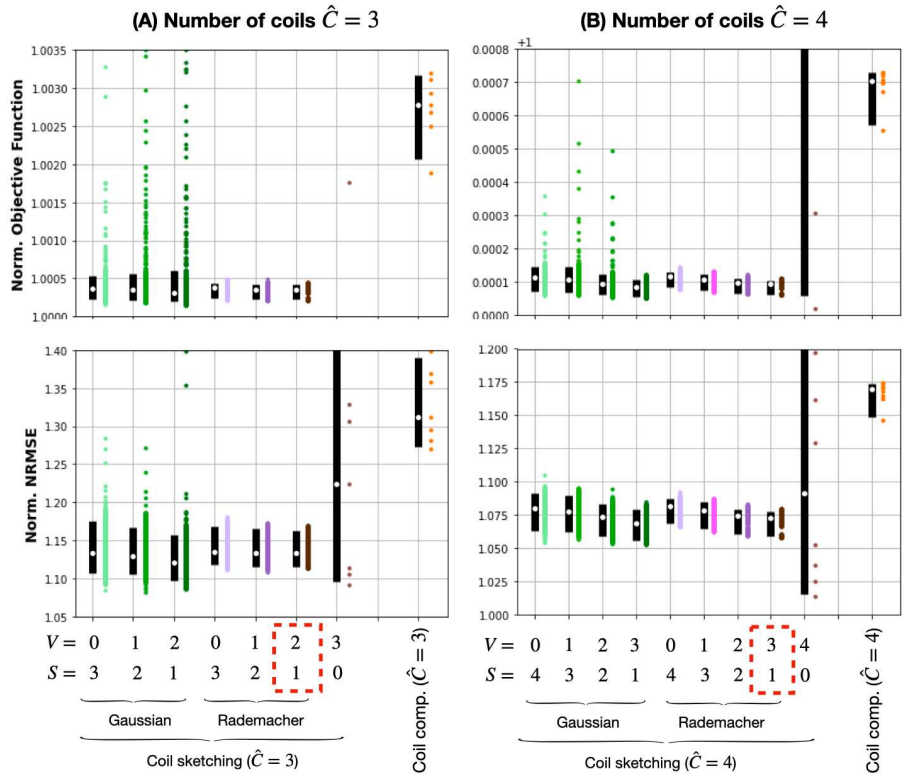}}
\caption[Sketching matrix design - 2D $\ell_1$-Wavelets reconstruction]{Evaluation of sketching matrix design through normalized objective function and normalized NRMSE for $\ell_1$-Wavelets reconstruction with 2D radial dataset. We report results  for reduced number of coils \textbf{(A)} $\hat{C}=3$ and \textbf{(B)} $\hat{C}=4$. The metrics are normalized by the value obtained in the baseline reconstruction with $C=8$ coils. The white circle and the black lines represent the median and 95\% confidence intervals respectively. Next to them, the colored circles represent the individual metric values for each slice and random draw of the sketching matrix. Rademacher distribution shows increased precision when compared to Gaussian distribution. Additionally, increasing the number of virtual coils $V$ increases reconstruction accuracy. However, when no sketched coils are present ($V=\hat{C}$), the optimization diverges, which suggests that in this case the sketched subproblems are not accurate approximations of the original problem
}
\label{fig2}
\end{figure*}

\begin{figure*}[h]
\centerline{\includegraphics[width=1.37\columnwidth]{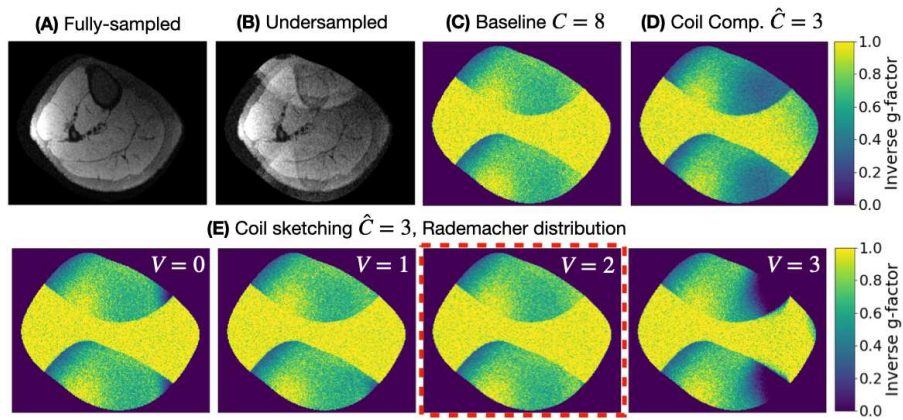}}
\caption[Sketching matrix design - 2D $\ell_2$ inverse g-factor analysis]{Evaluation of sketching matrix design through inverse g-factor for $\ell_2$-regularized reconstruction with 2D Cartesian dataset. \textbf{(C)} The baseline method ($C=8$) shows decreased SNR in the areas affected by aliasing and \textbf{(D)} aggressive coil compression ($\hat{C}=3$) shows additional SNR loss. \textbf{(E)} In coil sketching ($\hat{C}=3$) with Rademacher distribution, SNR improves as $V$ increases, and when $V=2=\hat{C}-1$, the inverse g-factor is virtually equivalent to the baseline method. %When $V=\hat{C}$, the inverse g-factor map worsens. 
Supporting Figure S3 shows exemplary reconstructed images.
}
\label{fig3}
\end{figure*}

The reconstruction times median and 95\% Confidence Intervals (95\%-CI) for coil compression $\hat{C}=8$ (baseline), $\hat{C}=3$, and $\hat{C}=4$ were 550 ms [506 ms, 592 ms], 425 ms [419 ms, 442 ms], and 442 ms [412 ms, 483 ms] respectively.  The times for coil sketching $\hat{C}=3$ and $\hat{C}=4$ were 440 ms [429 ms, 452 ms] and 449 ms [438 ms, 461 ms] respectively. 

In the same order, the GPU memory usages for coil compression were 371 MB, 339 MB, and 341 MB; and for coil sketching 344 MB and 348 MB.

Figure 3 presents an exemplary result of the second sketching matrix design experiment, the inverse g-factor assessment. Similarly, the baseline considered $C=8$ and the rest $\hat{C}=3$. The baseline (Figure 3C) shows the SNR decrement in the areas affected by aliasing, while aggressive coil compression $\hat{C}=3$ shows increased SNR loss. Figure 3E shows the inverse g-factor maps for coil sketching $\hat{C}=3$ with Rademacher distribution. SNR improves as we increase $V$. Furthermore, for $V=2$, the map is virtually equivalent to the baseline $C=8$. However, not including sketched coils ($V=\hat{C}=3$, $S=0$) penalizes SNR. %Supporting Figure 2 shows corresponding reconstructed images.

The reconstruction times median and 95\%-CI for coil compression $\hat{C}=8$ (baseline), coil compression $\hat{C}=3$, and coil sketching $\hat{C}=3$ were 236 ms [231 ms, 241 ms], 133 ms [130 ms, 136 ms], and 155 ms [154 ms, 158 ms] respectively. The GPU memory usages were 330 MB, 300 MB, and 308 MB respectively.

From these experiments, we conclude that including virtual coils in the sketching matrix ($V \geq 1$) increases robustness. However, at least one sketched coil must be included ($V < \hat{C}$, $S \geq 1$), otherwise the reconstruction becomes unstable. For the rest of the experiments, we set $V=\hat{C}-1$ and $S = 1$ for the coil sketching reconstructions. Furthermore, the Rademacher distribution presents higher precision when compared to Gaussian. Finally, coil sketching can potentially improve computational efficiency even in relatively small 2D datasets.

\subsection{3D non-Cartesian imaging}
\subsubsection{3D $\ell_1$-Wavelets Reconstruction}

Figure 4 shows the convergence analysis of the first 3D cones experiment using $\ell_1$-Wavelets reconstruction. We report the results for a reduced number of coils $\hat{C}=4$. 
% The baseline reconstruction uses $C=20$ and the rest of the methods use a reduced number of coils $\hat{C}=4$. 
For coil sketching, the sketching matrix considers Rademacher distribution and $V = \hat{C} - 1 = 3$, $S=1$. Figure 4A presents the convergence curves measured by the normalized distance to $\mathbf{x}^\infty$, Figure 4B shows the reconstructed images at $t=\infty$ for the three fastest methods, and Figure 4C further details convergence progression with the difference images with respect to $\mathbf{x}^\infty$ at three time points.

From Figure 4A, aggressive coil compression, $\hat{C}=4$, is the fastest to converge, followed by coil sketching, AccProxSGD, and baseline in that order. However, in Figure 4B, aggressive coil compression shows considerable image artifacts (arrow) due to the aforementioned energy loss, whereas coil sketching yields virtually the same reconstructed image as baseline. Thus, coil sketching is the fastest method that preserves image quality. 
From Figure 4C, at $t=110$s, baseline presents structural differences and considerable noise. Aggressive coil compression is close to convergence but shows structural differences in the area affected by the artifacts (lower right). AccProxSGD presents still some structural differences. Coil sketching shows similar results to aggressive coil compression but with reduced structural difference in the lower right region. At $t=150$s, the baseline reconstruction has increased noise, whereas coil sketching has virtually converged. %Supplemental Figure X shows these results for SGD as well. 

Figure 5 shows an example axial slice from the reconstructed volumes that illustrates the shading artifacts of aggressive coil compression. Visually, the baseline method, AccProxSGD, and coil sketching yield virtually the same reconstructed image.

Table \ref{table2} summarizes all measurements. Aggressive coil compression has the fastest convergence time (92s), followed by coil sketching (114s), and AccProxSGD (137s). Baseline is considerably slower (193s). However, aggressive coil compression has considerably worse image quality (NRMSE 0.12, SSIM 0.69, HFEN 0.16). Baseline and coil sketching have virtually equivalent quality metrics (NRMSE 0.09 vs. 0.09, SSIM 0.80 vs. 0.81, HFEN 0.10 vs. 0.10). AccProxSGD presents slightly worse quality metrics (NRMSE 0.10, SSIM 0.78, HFEN 0.10). Although, in theory, the baseline reconstruction and coil sketching should converge to the same solution, in practice, the images are not identical. The difference is explained by numerical differences due to the iterative solver and early termination (we terminate when updates are no longer noticeable visually).

\begin{table}[!h]
\begin{tabular}{|p{64pt}|p{30pt}|p{30pt}|p{30pt}|p{30pt}|}
\hline
& Base-line & Coil Comp. & Acc ProxSGD & Coil sketching \\
\hline
\rowcolor{gray!25}Duration (s) & 193 & 92 & 137 & 114 \\
Memory (GB) & 10.7 & 7.8 & 7.8 & 7.8   \\
\rowcolor{gray!25}N$^\circ$ FFTs & 520 & 168 & 192 & 240\\
N$^\circ$ Wav. & 13 & 21 & 24 & 18 \\
\rowcolor{gray!25}NRMSE & 0.09 & 0.12 & 0.10 & 0.09 \\
SSIM & 0.80 & 0.69 & 0.78 & 0.81\\
\rowcolor{gray!25} HFEN & 0.10 & 0.16 & 0.10 & 0.10 \\
\hline
\end{tabular}
\caption[Summary of 3D Cones $\ell_1$-Wavelets reconstruction]{Summary of 3D Cones $\ell_1$-Wavelets reconstruction. We benchmark Coil sketching through computational efficiency and image quality metrics.}
\label{table2}
\end{table}

\begin{figure*}[h]
\centerline{\includegraphics[width=1.40\columnwidth]{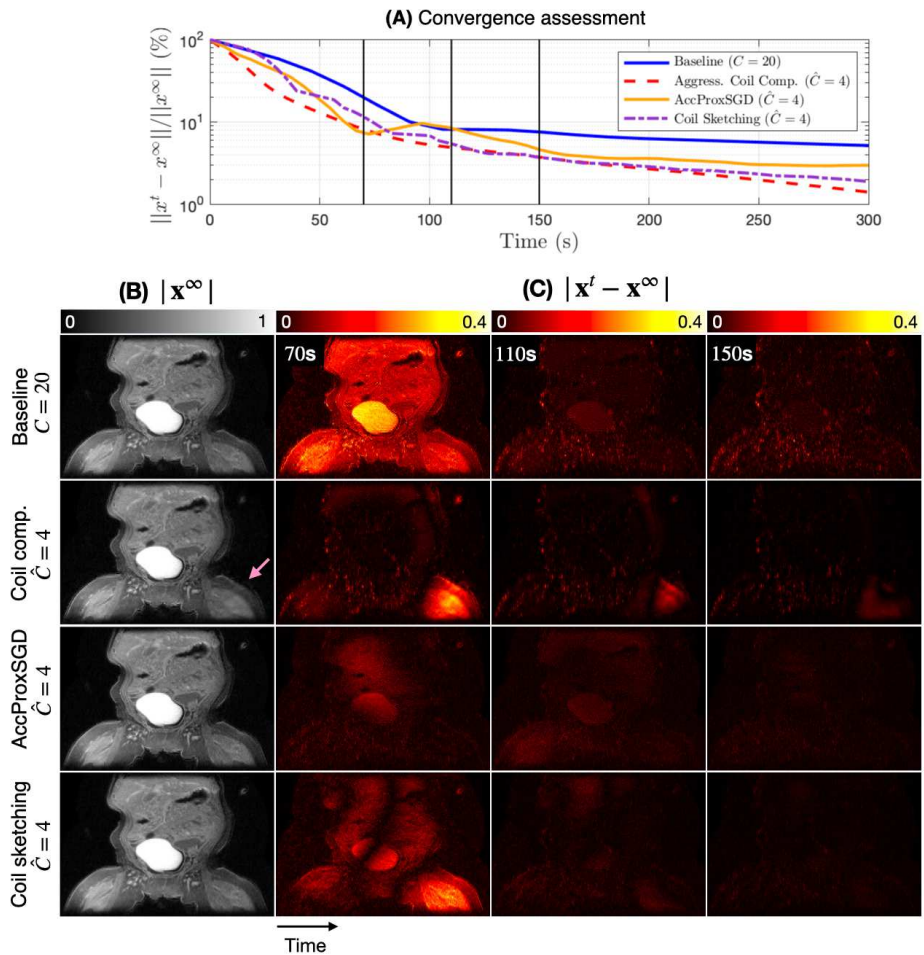}}
\caption[3D Cones $\ell_1$-Wavelets reconstruction - convergence evaluation]{$\ell_1$-Wavelets reconstruction with 3D cones. \textbf{(A)} We assess convergence speed measuring distance to $\mathbf{x}^\infty$. Aggressive coil compression is the fastest, closely followed by coil sketching. \textbf{(B)} However, $\mathbf{x}^\infty$ of aggressive coil compression presents shading artifacts (arrow) that compromise image quality. \textbf{(C)} Additionally, we assess $|\mathbf{x}^t - \mathbf{x}^\infty|$. At 110s, baseline has still some structural differences and noise, whereas the rest have almost converged. }

\label{fig4}
\end{figure*}

\begin{figure*}[h]
\centerline{\includegraphics[width=1.4\columnwidth]{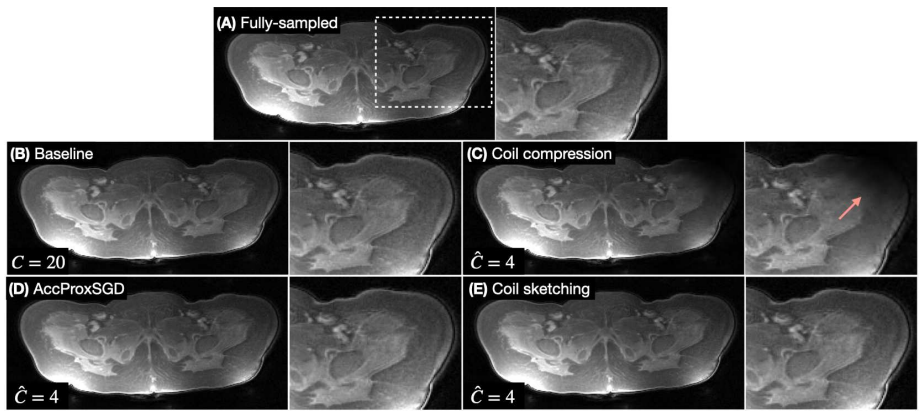}}
\caption[3D Cones $\ell_1$-Wavelets reconstruction - image quality visual evaluation]{Image quality visual evaluation for $\ell_1$-Wavelets reconstruction with 3D cones. We show an exemplary axial slice and compare the results to the \textbf{(A)} fully-sampled reference. \textbf{(C)} Aggressive coil compression presents considerable shading artifacts, whereas \textbf{(D)} AccProxSGD and \textbf{(E)} coil sketching presents practically equivalent quality as \textbf{(B)} baseline.}
\label{fig5}
\end{figure*}

The baseline reconstruction is significantly slower due to the number of coils, $C=20$, which is evidenced by the number of FFTs performed (520 versus $\approx$200 for the other methods). Additionally, coil sketching requires more FFTs than aggressive coil compression and AccProxSGD (240 versus 168 and 192 respectively). This difference is mainly due to the true gradient calculation (Algorithm 1), where we must use the original forward model, $\mathbf{A}$, with all $C=20$ coils. Nevertheless, once the subproblem is formed, we consider only the reduced number of coils, $\hat{C}$. Finally, although AccProxSGD performs fewer FFTs than coil sketching, the reconstruction is slower because it must load/unload to GPU memory the reduced number of coils $\hat{C}$ at each iteration, whereas coil sketching loads/unloads them only each time a new sketched subproblem is formed. Supporting Figure S4 shows a time profile that details the numerical operations and GPU memory transfers required for each tested algorithm.

\subsubsection{3D $\ell_1$-TV Reconstruction}
\label{subsubsec:res_L1TV}

Figure 6 shows the convergence analysis for the second 3D cones experiment with $\ell_1$-TV reconstruction. Similarly, we report the results for a reduced number of coils, $\hat{C}=4$. For coil sketching, the sketching matrix uses the Rademacher distribution, $V = \hat{C} - 1 = 3$, and $S=1$ as well.
Figure 6A presents the convergence curves, Figure 6B shows $\mathbf{x}^\infty$ for the three fastest methods, and Figure 6C difference images to $\mathbf{x}^\infty$.

From Figure 6A, coil sketching is the fastest method followed by the baseline reconstruction, AccProxSGD, and aggressive coil compression in that order. From Figure 6B, AccProxSGD presents blurring, which compromises image quality, whereas coil sketching and the baseline method yield virtually equivalent image quality. Aggressive coil compression is significantly slower. Thus, in this case, coil sketching is the fastest method and also preserves image quality. From Figure 6C, coil sketching and AccProxSGD are close to convergence at 1000s, whereas the baseline method still has significant structural differences. At 2000s, coil sketching has virtually converged, whereas AccProxSGD has slowed down and has similar structural differences than the baseline method.

Figure 7 shows an exemplary reconstructed axial slices, which further illustrates the significant image quality loss of AccProxSGD. Likewise, aggressive coil compression presents significant shading artifacts. Coil sketching and baseline yield virtually equivalent reconstructed images.

Table \ref{table3} summarizes all measurements. Coil sketching has the fastest convergence time (1,249s), followed by baseline (3,499s), AccProxSGD (3,829s) and aggressive coil compression (5800s). Regarding image quality metrics, coil sketching presents image quality virtually equivalent to the baseline reconstruction (NRMSE 0.11 versus 0.11, and SSIM 0.94 versus 0.92, HFEN 0.17 versus 0.17, whereas AccProxSGD has lower image quality (NRMSE 0.15, SSIM 0.77, HFEN 0.28). 

For this case, coil sketching is significantly faster than the baseline method, mainly, due to the different characteristics of the PDHG solver. Namely, in the data consistency step, PDHG solves an $\ell_2$-regularized least squares problem using CG at each iteration. This step requires a significantly larger number of forward model (or sketched forward model) passes at each iteration and, thus, when fewer coils are used, the number of FFTs decreases significantly more.

\begin{table}[h]
\begin{tabular}{|p{64pt}|p{30pt}|p{30pt}|p{30pt}|p{30pt}|}
\hline
& Base-line & Coil Comp. & Acc ProxSGD & Coil sketching \\
\hline
\rowcolor{gray!25}Duration (s)   & 3,499  & 5,740  & 3,829 & 1,249 \\
Memory (GB)    & 15.5   & 14.2   & 12.4 & 13.8   \\
\rowcolor{gray!25}N$^\circ$ FFTs & 7,600  & 11,440 & 5,432 & 5,600\\
N$^\circ$ Wav. & 19     & 66     & 6,790 & 42 \\
\rowcolor{gray!25}NRMSE          & 0.11   & 0.34   & 0.15 & 0.11 \\
SSIM           & 0.92   & 0.55   & 0.77 & 0.94\\
\rowcolor{gray!25}HFEN    & 0.17   & 0.47   & 0.28 & 0.17 \\
\hline
\end{tabular}
\caption[Summary of 3D Cones $\ell_1$-TV reconstruction]{Summary of 3D Cones $\ell_1$-TV reconstruction. We benchmark Coil sketching through computational efficiency and image quality metrics.}
\label{table3}
\end{table}

\section{Discussion}\label{sec5}

We have presented coil sketching as a novel, general algorithm for computationally-efficient MR image reconstruction. Coil sketching adapts theory from randomized sketching to the MR image reconstruction problem and leverages literature from both fields. Through multiple experiments, we have demonstrated that coil sketching improves computational efficiency by speeding up reconstruction in both small and large datasets without penalty on image quality, and that our method is general enough to be applicable to nearly all reconstruction formulations.

\begin{figure*}[h]
\centerline{\includegraphics[width=1.2\columnwidth]{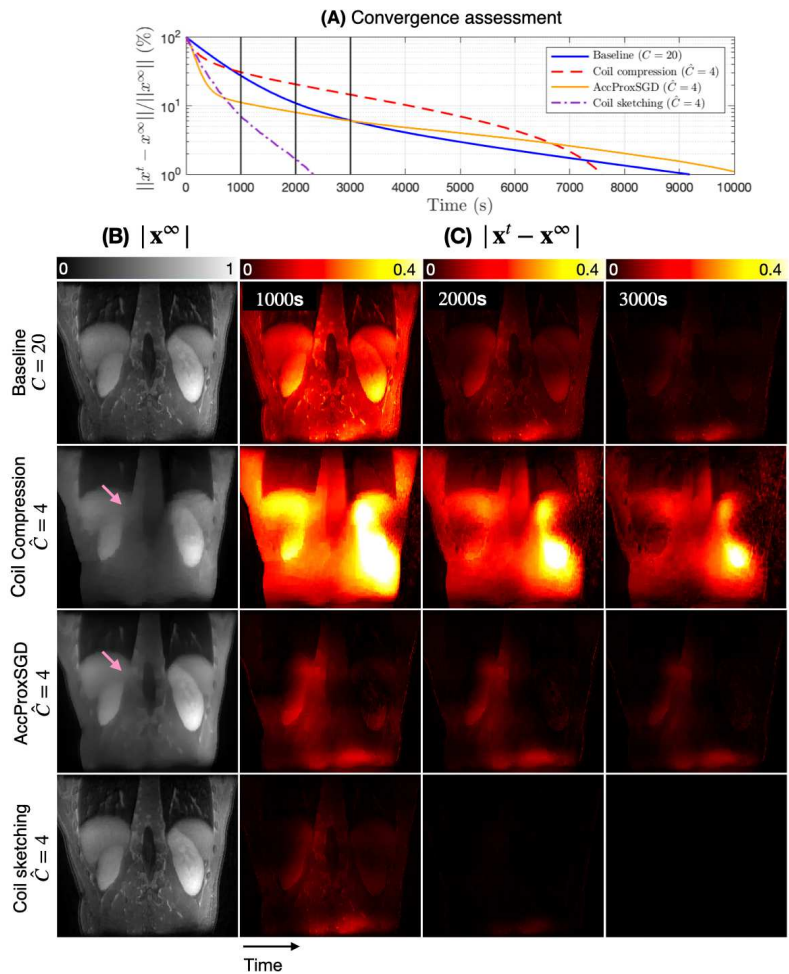}}
\caption[3D cones $\ell_1$-TV reconstruction - convergence evaluation]{$\ell_1$-TV reconstruction with 3D cones. \textbf{(A)} We assess convergence speed measuring distance to $\mathbf{x}^\infty$. Coil sketching is the fastest and \textbf{(B)} $\mathbf{x}^\infty$ shows virtually the same quality as baseline. \textbf{(C)} Additionally, we assess $|\mathbf{x}^t - \mathbf{x}^\infty|$. At 2000s, coil sketching has practically converged, whereas the rest of methods have still considerably structural differences.}
\label{fig6}
\end{figure*}

\begin{figure*}[h]
\centerline{\includegraphics[width=1.15\columnwidth]{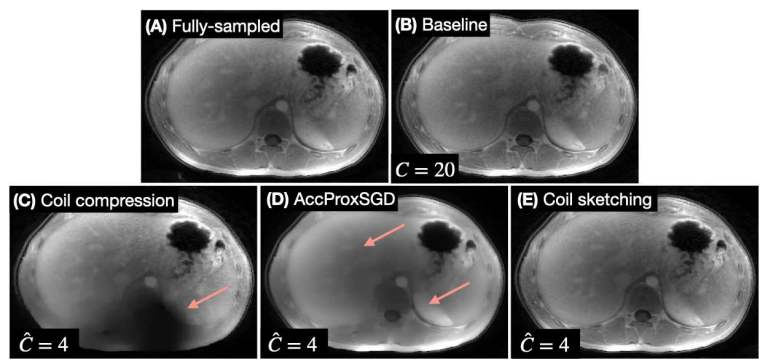}}
\caption[3D cones $\ell_1$-TV reconstruction - image quality visual evaluation]{Image quality visual evaluation for $\ell_1$-TV reconstruction with 3D cones. We show an exemplary axial slice and compare the results to the \textbf{(A)} fully-sampled reference. \textbf{(C)} Aggressive coil compression presents obvious shading artifacts, and \textbf{(D)} AccProxSGD presents some blurring. \textbf{(D)} Coil sketching yields virtually equivalent quality as \textbf{(B)} baseline. Blurring in AccProxSGD image evidences under-optimization. In iterative reconstruction, the low frequency components are recovered first, whereas the high frequency components are recover in the later stages. Blurring could be mitigated with a larger amount of iterations. However, the reconstruction time is already significantly longer than with coil sketching.}
\label{fig7}
\end{figure*}

Our method achieves faster reconstructions because of two main features. First, our structured sketching matrix lowers the computational cost of the forward model by reducing the number of coils effectively used during reconstruction. The reduced coil dimension yields faster iterations due to fewer computations (FFTs) and, more importantly, fewer GPU memory transfers in large datasets. 

Second, the novel sketched formulation of the reconstruction problem has advantageous convergence properties. The iterative sketch guarantees convergence to the original solution, which in practice yielded reconstructed images with equivalent image quality. Furthermore, the iterative sketch has better condition number dependence than SGD, which in our experiments resulted in fewer iterations required for convergence.

The first two experiments carefully designed and validated the structured sketching matrix. We aimed to leverage simultaneously randomized sketching theory, the Fourier structure of the MRI problem, and previous research on coil compression. The key idea was a structured design that produced linear combinations in the coil dimension and reduced the number of coils, $\hat{C}$, effectively used during reconstruction. We tested designs with different combinations of number of high-energy virtual coils ($V$) and sketched low-energy virtual coils ($S$). The best performing design for a sketching matrix of $\hat{C}$ coils included up to $\hat{C}-1$ high-energy virtual coils and one sketched coil. Additionally, these experiments showed that coil sketching can speed up reconstructions even in relatively small datasets, while preserving SNR as measured by the inverse g-factor.

Then, we evaluated the performance of coil sketching in our large-scale application of interest, 3D cones datasets. Coil sketching was the fastest method that did not compromise  image quality. In the $\ell_1$-Wavelets experiment, aggressive coil compression was faster, but it yielded significantly reduced image quality due to considerable energy loss. In contrast, coil sketching was slightly slower, but it yielded virtually the same image quality with 1.7x faster reconstruction speed when compared to the baseline reconstruction. Another disadvantage of aggressive coil compression is the need for $\lambda$ hyperparameter re-tuning, whereas coil sketching can re-utilize the same $\lambda$ as the baseline method since the formulation leverages information/energy from all $C$ coils via the true gradient.

% Similarly, coil sketching outperformed the previously proposed SGD. Although they have similar convergence rates per iteration [ref], SGD loads/unloads a different set of coils in GPU memory each iteration, which increases the time per iteration. In contrast, coil sketching only loads/unloads sets of coils when forming each sketched subproblem, a process that is repeated one order of magnitude less than the total number of iterations.
Similarly, coil sketching outperformed both the baseline reconstruction and the previously proposed SGD by leveraging both the reduced number of computations and increased memory efficiency. First, our method reduces the number of computations as suggested by the total number of FFTs, the most expensive operation in the forward model. Although coil sketching solves multiple sketched sub-problems and, thus, performs more iterations, our method requires less FFTs to achieve convergence. Table \ref{table2} shows that coil sketching performs less than half the number of FFTs than the baseline method.

Second, coil sketching benefits from the increased memory efficiency. 
The reduced forward model allows to solve the sketched sub-problems entirely in GPU with only one memory transfer. Coil sketching only loads/unloads coils when forming each sketched sub-problem, which occurs  significantly less times than the total number of iterations. As a comparison, although SGD performed a similar number of FFTs, SGD was significantly slower.
While both methods used the same number of coils $\hat{C}$, SGD loads/unloads a different set to GPU memory at each iteration to maintain an unbiased estimate of the gradient, which increases reconstruction time, as shown in the time profile (Supporting Figure S4).

Another important advantage of coil sketching is versatility. The  sketched subproblems can be solved using the same optimizer as in the original reconstruction problem, which makes coil sketching readily applicable to almost all types of image reconstructions (e.g. low-rank, dictionary learning). We evaluated this trait with the $\ell_1$-TV experiment, where TV regularization is not computationally-efficeint \cite{kamilov2016parallel}.

In the $\ell_1$-TV experiment, coil sketching greatly outperformed the other methods, achieving an approximately 2.8x faster reconstruction than the baseline method. The larger difference is due to the larger computational cost of the data consistency. Therein, PDHG solves an $\ell_2$-regularized least squares problem at each iteration, which requires performing multiple forward model (or sketched forward model) passes. Further details are shown in Section S2 of Supporting Material. Therefore, considering fewer coils ($\hat{C} < C $) reduces computational cost more significantly. We can also conclude that coil sketching is potentially more useful as the data consistency step has larger computational cost, which is the case in the massive multi-dimensional datasets such as DCE, 4D-flow or MRF.

Furthermore, SGD performed significantly worse as evidenced by the longer reconstruction time and decreased image quality (blurriness). First, in our implementation, SGD required a nested optimization algorithm to solve the proximal step of TV regularization, which increased the time per iteration. Second, in theory, SGD methods have worse condition number dependence than iterative sketching \cite{bubeck2015convex, pilanci2016iterative}, which could explain the significantly larger amount of iterations. Furthermore, image blurriness evidences under-optimization and the need for even a larger amount of iterations to recover the high frequency components.

In the general setting, randomized sketching reduces both computation time of the solution and memory footprint. Herein, coil sketching did not reduce memory footprint significantly when compared to the baseline method because the latter performed coil batching, i.e. loading/unloading the coils to memory by batches. While coil batching slows down computation time, this method was necessary due to GPU memory constraints.

Finally, although coil sketching has proven to be a useful method to reduce computational cost of image reconstruction, the method has limitations. First, the method improves reconstruction efficiency by reducing the computational cost of the data consistency step. Therefore, in cases where this step is not significantly more costly than the regularization step, such as the 2D cases, coil sketching savings are less significant. Second, coil sketching has a larger amount of hyperparameters to tune, including the number of sketched subproblems and the number of inner iterations at each sketched subproblem.

% \textcolor{Green}{
% \section{Conclusions}\label{sec6}
% }

\section{Conclusions} \label{sec6}
We have presented coil sketching, a novel, general algorithm for computationally-efficient MR image iterative reconstruction. The method synergistically leverages literature from randomized sketching, the Fourier structure of the problem, and coil compression via a carefully design structured sketching matrix that reduces the number of coils actively used during reconstruction.

% In this article, we have presented a novel, general algorithm for computationally-efficient MR image reconstruction. Coil sketching adapts theory from randomized sketching to the MR reconstruction problem by leveraging the Fourier structure and previous research on coil compression. Through multiple experiments, we have demonstrated the use of coil sketching to improve computational efficiency of reconstruction by speeding up the process without penalty on memory use nor image quality.

Future work will extend coil sketching to deep learning-based reconstruction, particularly to the state-of-the-art unrolled networks \cite{hammernik2018learning,mardani2018neural,aggarwal2019}, and to problems where both the sensitivity maps and images are optimized simultaneously \cite{ying2007jsense,holme2019enlive,dwork2020calibrationless}.

Finally, we are aware of the vast bibliography of stochastic methods similar to SGD \cite{xiao2014proximal,nitanda2014stochastic,pham2020proxsarah,chambolle2018stochastic, strohmer2009kazmarz} that could be adapted to the MRI. Future work aims to adapt them and perform a comprehensive comparison.

\vspace{0.5cm}
\section*{Acknowledgments}
We thank Ali B. Syed and Marcus T. Alley for technical support acquiring and accessing the datasets. We thank also John M. Pauly and Kawin Setsompop for the long discussions and suggestions that improved the work. Finally, we thank Matthew J. Middione and Gustavo Chau for their suggestions preparing the manuscript.

\section*{Data availability statement}
All the code for the experiments was implemented in Python using the Sigpy library \cite{ong2019sigpy}. In the spirit of reproducible research, the code and data necessary to reproduce most of the results in the paper is available at: 
\begin{verbatim}
https://github.com/julioscanoa/sketching_mri
\end{verbatim}

\section*{Supporting information}
The following supporting information is available as part of the online article:

\vskip\baselineskip\noindent
\textbf{Section S1.}
{Hessian sketch derivation }
\vskip\baselineskip\noindent
\textbf{Section S2.}
{Solvers implementation details}
\vskip\baselineskip\noindent
\textbf{Figure S1.}
{Sketching matrix design - 2D $\ell_1$-Wavelets reconstruction. HFEN and SSIM metrics. }
\vskip\baselineskip\noindent
\textbf{Figure S2.}
{Sketching matrix design - 2D $\ell_1$-Wavelets reconstruction. Reduced number of coils $\hat{C}=6$.}
\vskip\baselineskip\noindent
\textbf{Figure S3.}
{Exemplary reconstructed images from inverse g-factor experiment}
\vskip\baselineskip\noindent
\textbf{Figure S4.}
{Time profile of $\ell_1$-Wavelets reconstruction with 3D cones}

% \clearpage

\listoffigures
\listoftables
\vspace*{6pt}
\end{document}